\documentclass[aps,prd,twocolumn,amsmath,amssymb,floatfix,superscriptaddress]{revtex4-1}
\usepackage{graphicx}
\usepackage[usenames,dvipsnames]{xcolor}
\usepackage{braket}
\usepackage{hyperref}
\begin{document}
\title{Estimating the central charge from the R\' enyi entanglement entropy}
\author{A. Bazavov}
\affiliation{Department of Physics and Astronomy, The University of Iowa, Iowa City, IA 52242, USA}
\affiliation{Department of Physics and Astronomy, University of California, Riverside, CA 92521, USA}
\affiliation{Department of Computational Mathematics, Science and Engineering and
  Department of Physics and Astronomy, Michigan State University, East Lansing, Michigan 48824, USA}
\author{Y. Meurice}
\affiliation{Department of Physics and Astronomy, The University of Iowa, Iowa City, IA 52242, USA}
\author{S.-W. Tsai}
\affiliation{Department of Physics and Astronomy, University of California, Riverside, CA 92521, USA}
\author{J. Unmuth-Yockey}
\email[Contact Judah at ]{judah-unmuth-yockey@uiowa.edu}
\affiliation{Department of Physics and Astronomy, The University of Iowa, Iowa City, IA 52242, USA}
\email{judah-unmuth-yockey@uiowa.edu}
\author{Li-Ping Yang}
\affiliation{Department of Physics,Chongqing University, Chongqing 401331, China}
\author{Jin Zhang}
\affiliation{Department of Physics and Astronomy, University of California, Riverside, CA 92521, USA}
\definecolor{burnt}{cmyk}{0.2,0.8,1,0}
\def\lt{\lambda ^t}
\def\note{note}
\def\beq{\begin{equation}}
\def\enq{\end{equation}}
\newcommand{\Tr}{\text{Tr}}

\begin{abstract}
We calculate the von Neumann and R\'enyi bipartite entanglement entropy of the $O(2)$ model with a chemical potential on a 1+1 dimensional Euclidean lattice with open and periodic boundary conditions. We show that 
the Calabrese-Cardy conformal field theory predictions for the leading logarithmic scaling of these entropies are consistent with a central charge $c=1$. This scaling survives the time continuum limit and truncations of the microscopic degrees of freedom, modifications which allow us to connect the Lagrangian formulation 
to quantum Hamiltonians.
At half-filling, 
the forms of the subleading corrections imposed by conformal field theory allow the determination of the central 
charge with an accuracy better than two percent for moderately sized lattices.
We briefly  discuss the possibility of estimating the central charge using quantum simulators. \end{abstract}

\maketitle
\section{Introduction}
Conformal symmetry has been a major source of inspiration for theoretical physics during the last few decades \cite{1987gauge,difrancesco}. In two dimensions the conformal algebra is infinite dimensional and can be identified with the 
Virasoro algebra generating the reparameterization of the world sheet in string theory. This algebra admits central extensions 
labeled by the central charge, denoted $c$ hereafter. Known unitary representations with $c=1-6/(m(m+1))$ and $m=3,\dots ,6$ describe the 
critical behavior of the two-dimensional Ising and 3-states Potts models and their tricritical versions \cite{Belavin:1984vu,Friedan:1983xq,Dotsenko:1984if}.

In three dimensions, the possibility that a cusp on the boundary of the region of anomalous dimensions allowed by the conformal bootstrap corresponds to the Ising universality class  has triggered very interesting new developments
\cite{PhysRevD.86.025022,El-Showk2014,Kos2014}.
In four dimensions, the idea that electroweak symmetry breaking could result from new strong interactions at a multi-TeV scale with an approximate conformal symmetry \cite{PhysRevLett.77.1214, PhysRevD.55.5051,Chivukula:1996kg,PhysRevD.59.067702,Luty:2004ye,Dietrich:2005jn} protecting a light 
Higgs-Brout-Englert boson 
has motivated numerous lattice studies \cite{DeGrand:2010ba,Blum:2013mhx,Kuti:2015awa}.

QCD-like systems with
various numbers of fundamental fermions and also fermions in
different representations are being explored on the lattice
(the latest results for various models are presented, for instance,
in
Refs.~\cite{DeGrand:2015lna,Brower:2015owo,Arthur:2016dir,Appelquist:2016viq,Fodor:2016wal}).
Based on the Banks-Zaks argument, systems with a large number of
fermion flavors, $N_f$, feature a conformal phase~\cite{banks81}.
However,
precisely at what value of $N_f$ this happens
for a particular gauge group and fermion representation
remains a subject
of controversy. For the $SU(3)$ gauge group
and fermions in the fundamental representation,
some studies claim  observing conformal behavior at $N_f=12$ (see \cite{Cheng:2014jba} for instance)
while others (\emph{e.g.} Ref. \cite{Fodor:2016zil}) argue that $N_f=12$
is not conformal.

To probe the conformality, various lattice methods,
designed and well-tested for QCD, have been employed. The major
obstacle, however, is that large-$N_f$ theories are very different
from QCD. While in QCD the running of the coupling is fast enough
so that one can
probe both the ultraviolet perturbative and the infrared confining
phenomena, as manifested, for instance, in one of the basic and
extensively studied quantities, the static quark anti-quark
potential, large-$N_f$ theories require fine lattices and large volumes
to disentangle the physics from the lattice cutoff effects.
This makes identifying conformal theories from the massless extrapolations of massive lattice simulations a nontrivial task \cite{Hasenfratz:2011xn,Fodor:2011tu,Aoki:2012eq,zech2,Cheng:2014jba,Fodor:2016zil}. 

In the given examples, conformal symmetry is explicitly broken
by the lattice regularization and only re-emerges in a suitable continuum limit and infinite volume limit. 
Given the predictive power of conformal symmetries, it is important to identify the restoration of these symmetries in practical calculations at finite volumes.
The entanglement entropy may offer a promising direction in
understanding the conformal behavior of systems with finite-size
scaling and could be a more sensitive tool, especially for
small-size systems.
How the entanglement entropy of a subsystem scales with the its spatial volume provides useful information about
the symmetries present and the conformality of the phases of a model \cite{PhysRevLett.100.030504}. 
This is very well understood in two (one space and one Euclidean time)
dimensions where Calabrese and Cardy (CC) \cite{Calabrese:2004eu} have shown that various entanglement entropies scale like the logarithm of the size of the subsystem with coefficients proportional to the central charge. 

Calculations of the entanglement entropy
in lattice gauge theory with Monte Carlo methods have so far been
performed in pure gauge
theory~\cite{Velytsky:2008rs,Buividovich:2008kq,Itou:2015cyu} and two dimensional critical spin systems \cite{gliozziJHEP08}. 
Those calculations use the ``replica trick'' where $n$ sheeted Riemannian surfaces are glued together over an interval, but may require extra developments for theories with
fermions.  The entanglement spectrum for lower order R\'enyi entanglement entropies has been analyzed for fermion systems in three dimensions using determinantal Monte Carlo methods, with universal behavior found that could potentially be observed in cold-atom experiments \cite{porterprb16,drutpre16,drut1510}.
In the long run finite-size scaling of the 
entanglement entropy may provide a cleaner way to study conformal systems than just increasing lattice volumes and
decreasing lattice spacings in the hope of suppressing lattice
artifacts.

In the following, we use renormalization group (RG) based methods \cite{prd88,prd89,pre89,PhysRevLett.69.2863,PhysRevLett.75.3537}
to calculate the von Neumann and second order R\'enyi entanglement entropy of the classical $O(2)$ nonlinear sigma model with a chemical potential in 1+1 dimensions on a space-time lattice. This model is often used as an effective theory for the Bose-Hubbard model \cite{PhysRevB.40.546} and is good as a toy model for gauge theories in higher dimensions.  The model has vortex solutions, no long range order, and demonstrates a confinement-deconfienment transition of vortex-anti-vortex pairs.
This model \cite{PhysRevA.90.063603} has a superfluid (SF) phase where we expect to observe the CC scaling and multiple Mott insulator phase lobes lacking the CC scaling.

By using rectangular lattices of increasing spatial size and very large (Euclidean) temporal sizes we probe the zero temperature entanglement entropy. We focus on half-integer charge density where the entropies considered are extremal. We then take the time continuum limit and truncate the microscopic degrees of freedom in such a way that we obtain a Hamiltonian that can be quantum simulated \cite{PhysRevA.90.063603,PhysRevD.92.076003}. These modifications should not affect the universal parts of the scaling.  Our goal is to demonstrate that the constraints imposed  by conformal symmetry on the finite size scaling, as well as conjectures \cite{Fagotti-Calabrese:2011,PhysRevLett.104.095701,Calabrese-Essler:2010} explaining oscillations in the scaling, allows us to identify conformal behavior for modest lattice sizes.

The motivation for relating this model to a model that can be quantum simulated on optical lattices with cold atoms is prompted in current challenges with classical computation.  It would be not only valuable to have efficient calculational tools for understanding conformal behavior for more complex, higher dimensional systems, but also to completely overcome the difficulty with large volume, small lattice spacing calculations entirely.  This can be done by using quantum simulation, which can already reach volumes in 3+1 dimensions on the same order as classical computation and larger volumes are expected.  This idea is pursued in more detail in \cite{myfirstprl}.

Manipulations of small one-dimensional systems of cold atoms trapped in optical lattices have allowed experimental measurements of the second order R\'{e}nyi entanglement entropy \cite{2015arXiv150901160I} 
using a beam splitter method proposed in Ref.  \cite{PhysRevLett.109.020505}.
These measurements have been performed for small chains of four atoms. A more recent experiment on thermalization \cite{Kaufman2016} involves six atoms. 
It is expected that in the near future, manipulations of 
chains with twelve or more atoms will be possible \footnote{private communication, Philipp Preiss}.

The paper is organized as follows: in Section \ref{sec:renyi} we review the R\'{e}nyi entropy and the corresponding conventions used in this paper.  We also discuss the currently understood asymptotic scaling in the R\'{e}nyi entropy as a function of system size.  In Section \ref{sec:o2model} we introduce the $O(2)$ nonlinear sigma model on a lattice and the tensor formulation of the model.  We give explicit tensor elements and discuss the isotropic and anisotropic coupling limits used in this paper as well as some results obtained in those limits.  In Section \ref{sec:fits} we give results for fits to R\'{e}nyi entanglement entropy data. We consider the scaling of the entanglement entropy as a function of system size. We also go into detail about the methodology used in our fits and make comparisons with theoretical predictions.  In Section \ref{sec:qsim} we discuss the possibility of quantum simulating the $O(2)$ nonlinear sigma model and a possible quantum Hamiltonian that could be used for simulation.  We also consider finite temperature effects to the entanglement entropy.  Finally in Section \ref{sec:conclusion} we give concluding remarks about in what possible directions work could proceed, and other possible implications of this work.

\section{The R\'enyi entanglement entropy}
\label{sec:renyi}

For the calculation of the entanglement entropy, we will restrict ourselves to 1+1 dimensional space-time, or one space and one Euclidean time dimension, where the one dimensional space has an even number of sites.  For all results in this work the system was divided into two identical parts, each one half the size of the entire spatial dimension, $N_s$ (justification for this can be found in the supplemental material of Ref. \cite{myfirstprl}).  Other partitions of the system were considered as checks and exploratory.  Tracing over one of the halves, we obtain the reduced density matrix $\hat{\rho}_A$ for the other half (denoted $A$),
\begin{equation}
	\hat{\rho}_{A} = \Tr_{\text{env}}[\hat{\rho}]
\end{equation}
where the trace is over the ``environment'' leaving only the sub-system defined as $A$.  The $n$-th order R\'enyi entropy is defined as 
\begin{equation}
S_n(A)\equiv \frac{1}{1-n}\ln(\Tr[\hat{\rho}_{A}^{n}]) \ . 
\end{equation}
The limit $n\rightarrow 1^+$ is the von Neumann entanglement entropy, or the first order R\'{e}nyi entropy. 
$S_2$ is the second order R\'{e}nyi entropy, and was measured in recent cold atom experiments \cite{2015arXiv150901160I}.
An important goal for future work is to estimate the central charge, $c$, from 
empirical data. Using the transformation properties of the 
energy-momentum tensor and the Ward identities from CFT, CC established that, to leading order, the R\'{e}nyi entropy scales linearly with the logarithm of the spatial volume.  The constant of proportionality is the central charge multiplied by a rational that depends on the order of the R\'{e}nyi entanglement entropy and the boundary conditions: 
\begin{equation}
S_n(N_s) = 
\begin{cases}
	K_n+\frac{c(n+1)}{6n}\ln(N_s) & \text{for PBC} \\
    K_n'+\frac{c(n+1)}{12n}\ln(N_s) & \text{for OBC}.
    \label{eq:lcft}
\end{cases}
\end{equation}
The intercept is non-universal 
and different in the four situations considered here.

The calculation of $S_n$ can be performed \cite{PhysRevE.93.012138} using blocking (coarse-graining) methods \cite{prd88,prd89,pre89,PhysRevLett.69.2863}.  In this work we used the density matrix renormalization group (DMRG) with
matrix product states (MPS), as well as exact blocking formulas \cite{prd88,prd89,pre89,PhysRevA.90.063603} with the tensor renormalization group method (TRG), and the only approximation in these methods consists of truncating the number of states (called $D_{\text{bond}}$).
The errors associated with this truncation will be discussed later.

\section{The $O(2)$ model}
\label{sec:o2model}

In the following, we consider the classical $O(2)$ model on a  $N_s\times N_{\tau}$ rectangular lattice with sites labeled $(x,t)$. This is a generalization of the Ising model where the local spin is allowed to take values on a circle,  
making an angle $\theta$ with respect to some direction of reference. This angle can be interpreted as the phase of a complex field and the model has 
an exact charge conjugation symmetry, $\theta \rightarrow -\theta$, interchanging particles and anti-particles.
This symmetry can be broken by adding a chemical potential $\mu$ to the angle gradient 
 \cite{Hasenfratz:1983ba}. 
The partition function reads: 
\begin{equation}
Z = \frac{1}{2\pi} \int \prod_{(x,t)} d\theta_{(x,t)} {\rm e}^{-S}
\label{eq:bessel}
\end{equation}
with
 \begin{eqnarray}
      S=&-&  \beta_{\tau}\sum\limits_{(x,t)} \cos(\theta_{(x,t+1)} - \theta_{(x,t)}-i\mu)\cr&-&\beta_s\sum\limits_{(x,t)} \cos(\theta_{(x+1,t)} - \theta_{(x,t)}).
      \end{eqnarray}
We use periodic boundary conditions (PBC) or open boundary conditions (OBC) in space and always PBC in time. In the following, we define the charge density as
\begin{equation}
	\lambda \equiv \frac{1}{(N_s\times N_{\tau})} \frac{\partial \ln (Z)}{\partial \mu}.
\end{equation}
We will start with the situation where the relativistic interchangeability between space and time is present ($\beta_s=\beta_\tau$), as is typical in lattice gauge theory simulations. Later, we will take the time continuum limit and switch to the Hamiltonian formulation. 

For numerical purposes, and in order to connect the Hamiltonian formulation 
to quantum simulators, it is convenient to introduce discrete degrees of freedom on the links (bonds) of the lattice. 
Using Fourier expansions \cite{RevModPhys.52.453,Banerjee:2010kc,prd88}, one can show \cite{PhysRevA.90.063603,PhysRevE.93.012138} that 
the partition function can be expressed in terms  of a transfer matrix $Z=\Tr [\mathbb{T}^{N_\tau}]$ where 
the matrix elements of $\mathbb{T}$  have the explicit form
\begin{eqnarray}
&\ &\mathbb{T}_{(n_1,n_2,\dots, n_{{N_s}})(n_1',n_2',\dots, n_{N_s}')}=\cr &\ &\cr&\ &\sum_{\tilde{n}_{1}\tilde{n}_{2}\dots \tilde{n}_{N_s}} T^{(1,t)}_{\tilde{n}_{N_s}\tilde{n}_{1}n_1 n_1'}T^{(2,t)}_{\tilde{n}_{1}\tilde{n}_{2}n_2n_2'\dots }\cr &\ &\dots  T^{(N_s,t)}_{\tilde{n}_{N_{s}-1}\tilde{n}_{N_s}n_{N_s}n_{N_s}' },
\label{eq:tmspace}
\end{eqnarray}
with
\begin{eqnarray}
	T^{(x,t)}_{\tilde{n}_{x-1}\tilde{n}_x n_x n_x'} &=&  \sqrt{I_{n_{x}}(\beta_	\tau)I_{n'_x}(\beta_s)\exp(\mu(n_x+n_x'))}\nonumber \\
	& \ &\sqrt{I_{\tilde{n}_{x-1}}(\beta_s)I_{\tilde{n}_x}(\beta_\tau)} 	\delta_{\tilde{n}_{x-1}+n_x,\tilde{n}_x+n_x'} \ .
	\label{eq:tensor}
\end{eqnarray}
When $N_s$ is a power of 2, the traces in the spatial directions in Eq. \eqref{eq:tmspace} can be performed recursively and combined with a truncation of the number of states kept in the time direction \cite{PhysRevA.90.063603,PhysRevE.93.012138}. The accuracy of this tensor renormalization group method has been tested against sampling methods \cite{PhysRevE.93.012138}.

We can interpret $\mathbb{T}^{N_\tau}$ as a density matrix $\hat{\rho}$ if we normalize by the trace of the matrix.
It is important to understand that the classical spin model described above can be taken in a limiting form as a quantum model in one spatial dimension. In the following, we always take $N_\tau \gg N_s$ and extrapolate to infinite $N_{\tau}$. This corresponds to the zero temperature limit in the quantum terminology. 
Finite temperature effects will be discussed in Sec. \ref{sec:qsim} and were considered in Ref. \cite{PhysRevE.93.012138}.

The SF phase is characterized by a response of the charge density, $\lambda$, to a change in the chemical potential.  This is illustrated in Refs. \cite{Banerjee:2010kc,PhysRevA.90.063603}. In contrast, in the Mott phases, the 
charge density keeps a fixed integer value as we increase $\mu$. This lack of response is somewhat puzzling in the functional integral formulation and is often 
called the  ``Silver blaze" phenomenon \cite{PhysRevLett.91.222001} in the context of finite temperature QCD.  Another characterization of the two phases is by the scaling of the R\'{e}nyi entropy as a function of the volume of space.

In the following we focus on two cases for two different relationships between the spatial and temporal couplings.  We consider $\beta = 0.1$, $\mu \simeq 3$ (case 1), and $\beta = 2$, $\mu = 0$ (case 2).  Both of these situations are considered in the limits of isotropic coupling ($\beta_{s} = \beta_{\tau}$), and anisotropic coupling ($\beta_{s}\beta_{\tau} = \text{const.}$ with $\beta_{\tau} \rightarrow \infty$ and $a \rightarrow 0$, where $a$ is the temporal lattice spacing).  In case 1 the SF transition is driven by an increase in the chemical potential at fixed $\beta$.  This is the transition driven by fluctuations in density.  For case 2 the transition is driven by the presence of vortices and is the Berezinskii-Kosterlitz-Thouless (BKT) transition.  

\subsection{Isotropic Coupling}
In this section we consider the case where the coupling in space and time are the same.  This is a classical statistical two-dimensional spin system.  Using TRG we can block a (Euclidean) time slice of the lattice and consider it as a transfer matrix.  From it we can calculate a ``zero-temperature'' density matrix by taking $N_{\tau} \gg N_{s}$ (typically $N_{\tau} \approx 2^{20}N_{s}$ in practice, although even larger sizes may be used).  Then, from the density matrix one can make the reduced density matrix and calculate the R\'{e}nyi entropy of the desired order.

\begin{figure}
	\includegraphics[width=0.5\textwidth]{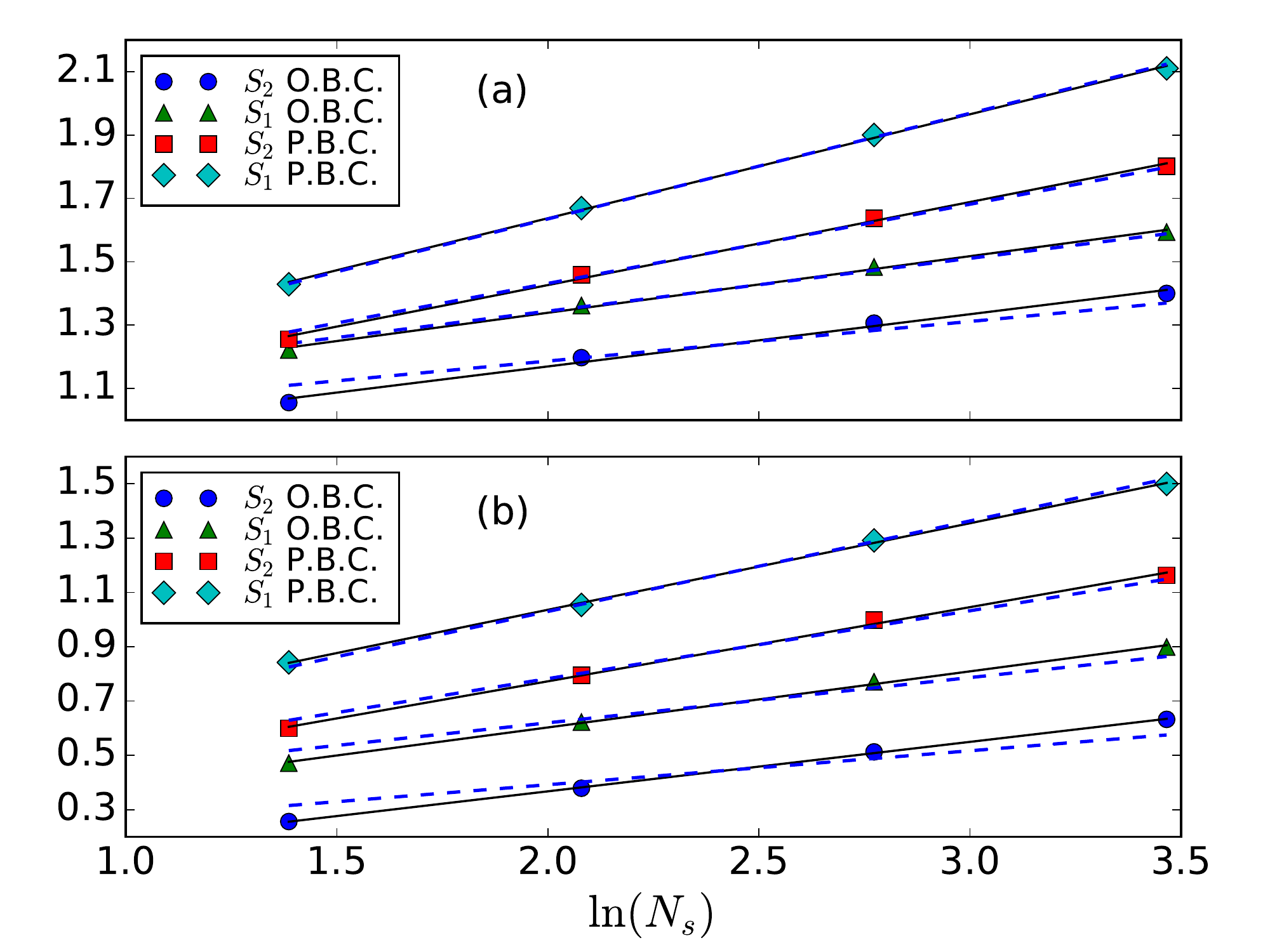}
\caption{(a) The second order ($S_{2}$) and first order R\'{e}nyi ($S_{1}$) entropies for case 1, $\mu =2.99$ and $\beta_{s} = \beta_{\tau} = 0.1$, both PBC and OBC.  The solid, black lines are linear fits to the data, and the dashed, blue (online) lines are fits of the intercept with  the CFT slopes. (b) Same quantities for case 2, $\mu=0$ and $\beta_s=\beta_\tau=2$.}
\label{fig:cc}
\end{figure}

The values of $S_1$ and $S_2$ are shown in Fig. \ref{fig:cc} for $N_s=4,8,16$, and $32$, for PBC and OBC.
These results are compared with the leading CFT prediction of Eq. \eqref{eq:lcft}
by just fitting the intercept with the CFT slope fixed.  For isotropic calculations the TRG calculations kept up to 250 states.  The figure shows that the discrepancies are rather small and most visible for $S_2$ with OBC for case 1. 
In case 2, the discrepancies are slightly more pronounced. In all cases, the discrepancies are due to subleading corrections not taken into account in the fits, rather than the small numerical errors.

\begin{figure}
    \includegraphics[width=0.5\textwidth]{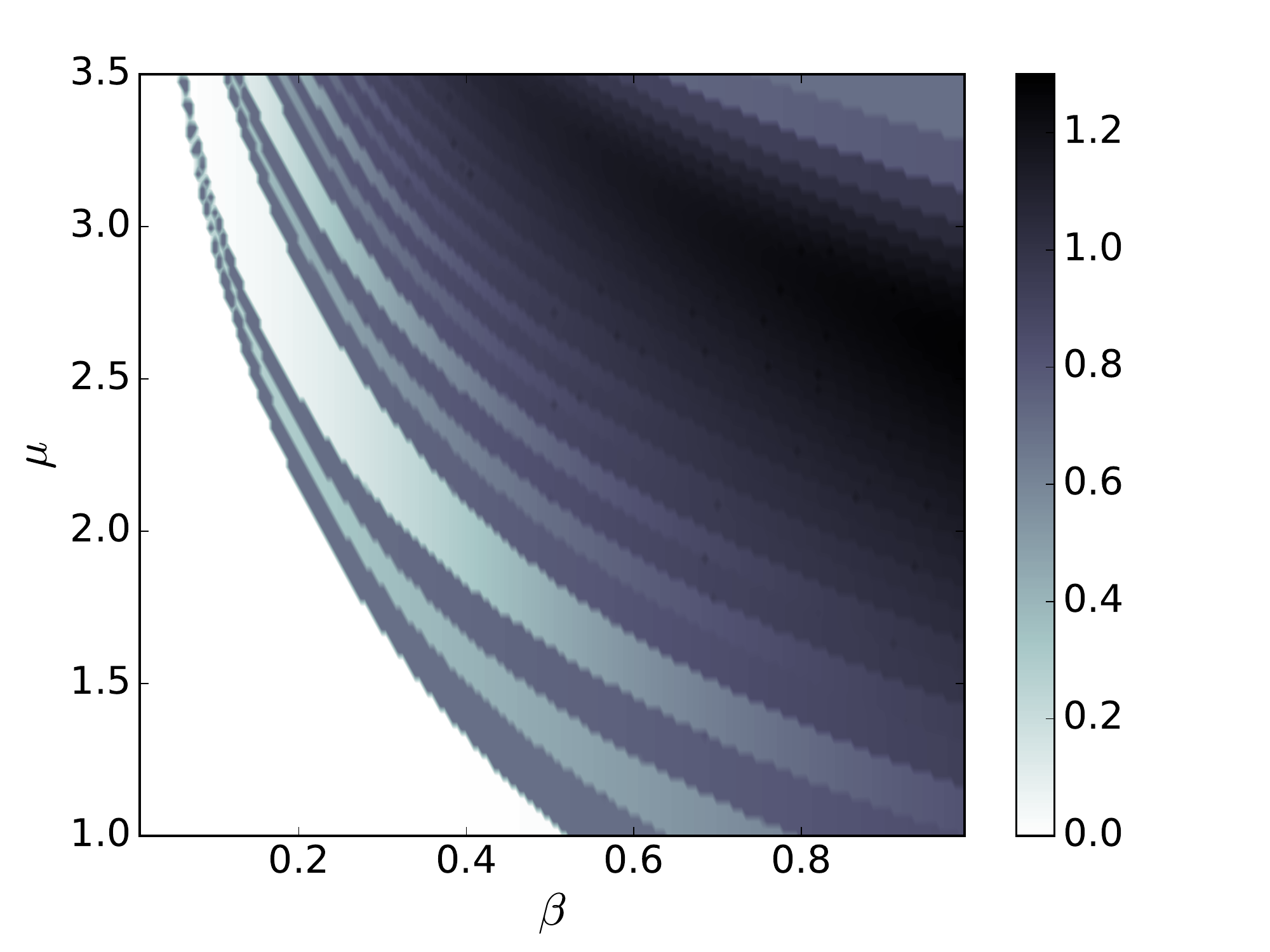}
	\caption{Second order R\'{e}nyi entropy in the $\beta$-$\mu$ plane for $N_{s}=4$ showing the various lobes of charge densities.  $\lambda = 0, 1,$ and $2$ are the prominent bright regions stemming from the $\beta=0$ axis.}
	\label{fig:ree}
\end{figure}

Fig. \ref{fig:ree} gives the values of $S_{2}$ for $N_{s} = 4$ across a region of the $\beta$-$\mu$ plane.
This figure shows lobes corresponding to a fixed charge density.  The largest and most prominent is the $\lambda = 0$ lobe, followed by much thinner $\lambda = 1$ lobe above.  The lobes continue as long as the truncation on the number of states used in the TRG calculation can support the charge density.  Fig. \ref{fig:ree} also shows 3 plateaus in the SF regions between each Mott lobe.  These plateaus were investigated and related to the charge density in the isotropic limit in \cite{PhysRevE.93.012138}.
In the next section we consider the time-continuum limit of the classical $O(2)$ model and how the phase diagrams transforms through taking that limit.

\subsection{Anisotropic Coupling}
We now proceed to take the time continuum limit. This can be achieved by taking $\beta_\tau$ very large while keeping constant the product $\beta_{s} \beta_{\tau}$
, and keeping $\mu \beta_{\tau}$
tuned to the desired charge density.
For case 1, the limit of the chemical potential must be done carefully in order to maintain a fixed charge density corresponding to half-filling.  For small volumes half-filling takes place around $\mu\beta_{\tau} = 0.5$ as $\beta_{\tau} \rightarrow \infty$, but not all the data collected for larger volumes was necessarily done at that parameter specification, and instead the parameters were tuned to maintain half-filling.
The time continuum limit in the tensor formulation defines \cite{PhysRevA.90.063603} a rotor Hamiltonian \cite{PhysRevD.17.2637,kogut79}:
\begin{equation}
	\hat{H}=\frac{U}{2}\sum_x \hat{L}_x^2-\tilde{\mu}\sum_x 		\hat{L}_x-2J\sum_{\left<xy\right>}\cos(\hat{\theta}_x-			\hat{\theta}_y) \ ,
\label{eq:rotor}
\end{equation}
with $[\hat{L}_x,{\rm e}^{i\hat{\theta} _y}]=\delta_{xy}{\rm e}^{i\hat{\theta} _y}$. It's possible to truncate to finite integer spin and approximate these commutation relations \cite{PhysRevA.90.063603}.
The normalization has been chosen in such a way that the coupling constants in the
Bose-Hubbard model used in Ref. \cite{2015arXiv150901160I}, and here in the $O(2)$ model are the same: $\beta_s \beta_\tau \equiv 2J/U$, and $\mu \beta_\tau \equiv \tilde{\mu}/U$.

\begin{figure}
\includegraphics[width=0.52\textwidth]{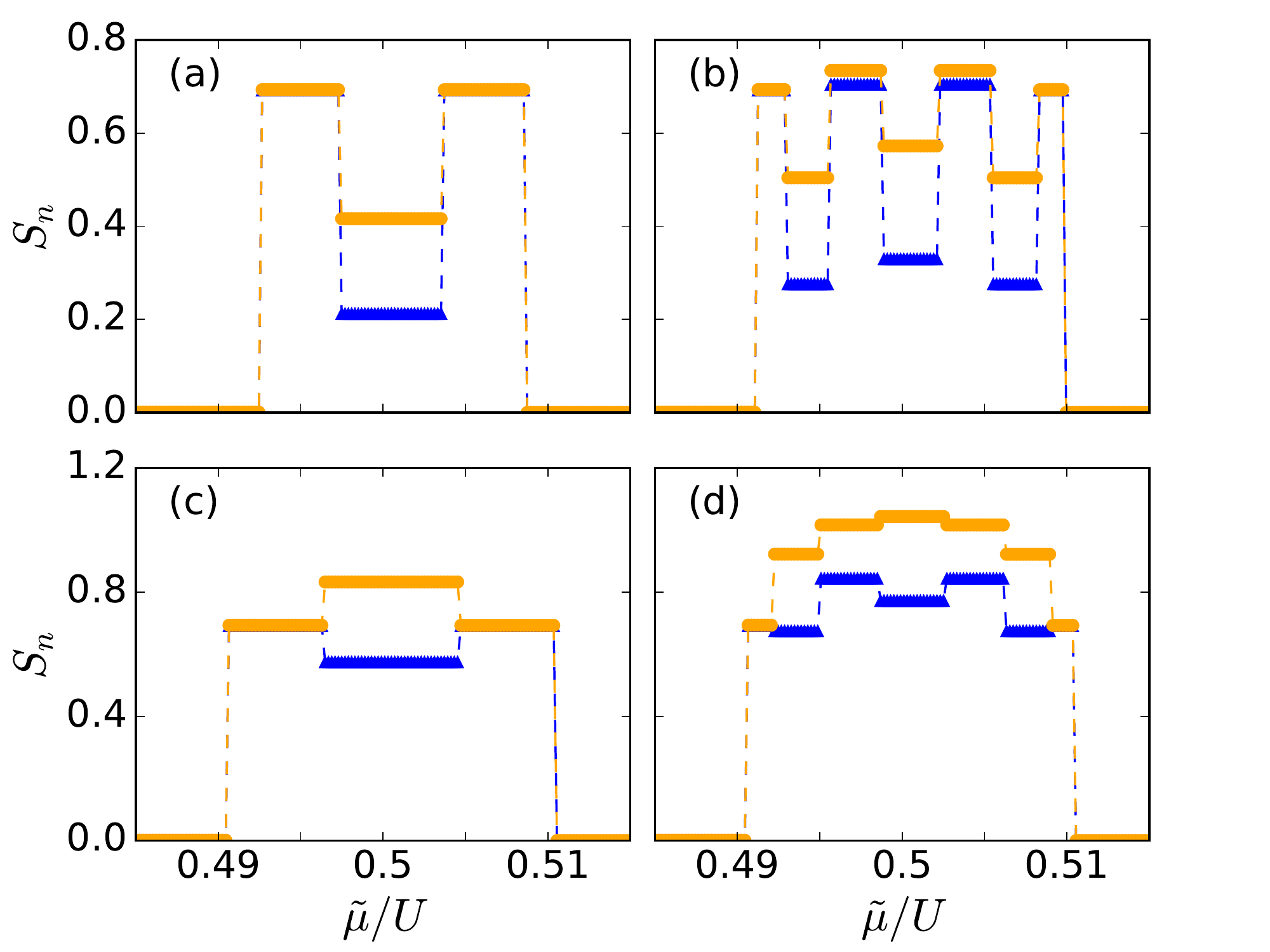}
\caption{The first order (orange online) and second (blue online) order R\'{e}nyi entropy as a function of $\tilde{\mu}/U$ in the region around case 1. (a), (b) Entropies for $N_{s} = 4$ and $N_{s} = 8$, respectively, with OBC.  (c), (d) Entropies for $N_{s} = 4$ and $N_{s} = 8$, respectively, with PBC.  The second R\'{e}nyi entropy maintains a local minimum around $\tilde{\mu}/U = 0.5$ for all cases, however the PBC data for the first R\'{e}nyi entropy has a maximum.  For this data $\beta_{\tau} = 500$ and $D_{\text{bond}} = 101$.}
\label{fig:entent_x4-8_sites}
\end{figure}

In the following we primarily use the spin-1 approximation which can also be implemented 
in the original isotropic formulation by setting the tensor elements in Eq. \eqref{eq:tensor} to zero for space and time tensor indices strictly larger than 1 in absolute value (so only 3 states remain).  
The Hamiltonian is then a spin-1 XY model with a chemical potential and an ion anisotropy. 
In addition, for 
large enough chemical potential, the $n=-1$ component decouples and we are approximately left with a spin-1/2 XY model. Furthermore, for $\tilde{\mu}=U/2\gg J$, there is an approximate connection with the Bose-Hubbard model 
\begin{equation}
	H=\frac{U}{2}\sum_x n_x(n_x-1)-J\sum_{x}(a^{\dagger}_xa_{x+1}+h.c.) .
    \label{eq:BH-ham}
\end{equation}
The Hamiltonian in Eq. \eqref{eq:rotor} is never explicitly used in the blocking procedure with TRG.  In practice, the TRG tensors used are the same, however the coupling constants that appear in the local tensor's definition are tuned to reflect the scenario under consideration.  Again, using the TRG the same way, one calculates the R\'{e}nyi entropy for an approximately zero-temperature situation.  The only minor change is that due to the increased coupling in the time direction, $N_{\tau}$ needs to be adjusted to compensate the smaller lattice spacing.  This means increasing $N_{\tau}$ even more, a facile task while using a blocking method.
We show slices of the R\'{e}nyi entropy in the region of case 1 for $N_{s} = 4$ and $8$, OBC, in Fig. \ref{fig:entent_x4-8_sites}.  A more extensive plot of $S_{2}$ for the $O(2)$ model in the time-continuum limit can be found in \cite{myfirstprl} for a large range of couplings along with a comparison to the Bose-Hubbard Model.

In order to check the TRG calculations of $S_n$ in the time continuum limit (Eq. \eqref{eq:rotor}), we have used DMRG \cite{PhysRevLett.69.2863} which has been used to calculate the ground state entanglement entropy and R\'{e}nyi entropy for similar Hamiltonians \cite{PhysRevB.92.041120, PhysRevB.84.085110, PhysRevLett.107.067202}.
Calculations with MPS optimization \cite{PhysRevLett.75.3537} have been performed using the ITensor C++ library \footnote{(version 2.7.10), http://itensor.org/}. 
We run enough sweeps for the entropy to converge to at least $10^{-8}$,  and a large number of states, up to $1500$, was kept so that the truncation error is less than $10^{-10}$. The comparison of the 
results with the two methods showed excellent agreement at small volume (typically 9 digits for $N_s=4$) but the discrepancies increased  with the volume (typically 3 digits agreement for $N_s=32$). We believe that the DMRG results are more accurate because firstly, it can keep many more states than the TRG by using sparse linear algebra libraries.  Secondly, the truncations are made step-by-step trying to maximize the entanglement entropy.  Finally, DMRG uses an environment sweep method which optimizes the ground state wave-function iteratively. 
For these reasons, we have used the DMRG results for the fits 
that follow. 

\section{Fits to $S_{n}$}
\label{sec:fits}

In this section we give some results for fits to the isotropic and anisotropic data, as well as for fits to the DMRG data.  The primary deviations from the leading order linear behavior of the R\'{e}nyi entropy come from finite volume effects and parity oscillations.  These deviations were found most prominently in OBC data, and the second order R\'{e}nyi entropy.  Oscillations were found between sizes $N_{s}$ mod 4 = 0 and $N_{s}$ mod 4 = 2, although in case 1 $S_{1}$ with PBC had none, as well as $S_{1}$ in case 2.  Corrections to the leading-order CFT behavior to account for these oscillations have been conjectured \cite{Fagotti-Calabrese:2011,PhysRevLett.104.095701,Calabrese-Essler:2010}, and we check their validity for the anisotropic $O(2)$ model.  In addition, non-oscillatory finite volume corrections have been derived \cite{Cardy-Calabrese:2010} and in the following we attempt to take all these corrections into account in order to fit the data as best as possible.

Initially, all the data was fit to the leading order CFT prediction (Eq. \eqref{eq:lcft}), with the slope and intercept as the two free parameters.  This was done for spatial volumes which matched the TRG blocking volumes, \emph{e.g.} $N_{s} = 2^{\ell}$, and since these sizes are multiples of 4, no oscillations were present.  These fits were done for both the DMRG and TRG data in both the isotropic and anisotropic limits.   The results for the slope fits are reported in Table \ref{table:naive-slopes} for all cases considered.

\begin{table}
  \begin{tabular}{||l|c|c|c|c||}
     \hline 
     case 1 & isotropic & anisotropic & DMRG & $c=1$ CFT \\
     \hline\hline
     $S_1$ PBC & 0.319 & 0.311 & 0.327 & $0.\bar{3}$ \\ \hline
     $S_2$ PBC & 0.273 & 0.265 & 0.267 & 0.25\\\hline   
     $S_1$ OBC & 0.207 & 0.208 & 0.195 & $0.1\bar{6}$ \\ \hline
     $S_2$ OBC & 0.182 & 0.152 & 0.168 & 0.125\\\hline   
     \hline  
     case 2 & isotropic & anisotropic & DMRG & $c=1$ CFT \\
     \hline\hline
	 $S_1$ PBC & 0.328 &  0.296 & 0.329 & $0.\bar{3}$ \\ \hline
     $S_2$ PBC & 0.262 &  0.229 & 0.250 & 0.25\\\hline   
     $S_1$ OBC & 0.179 &  0.152 & 0.159 & $0.1\bar{6}$ \\ \hline
     $S_2$ OBC & 0.165 &  0.148 & 0.140  & 0.125\\\hline   
       \end{tabular}
 \caption{Slopes of the R\'{e}nyi entropies using the leading-order linear fit.  The fits were done using the same volumes used in the TRG calculations: $N_{s} = 2^{\ell}$.  For data at these volumes oscillations do not appear; oscillations occur between volumes: $N_{s} \text{ mod } 4 = 0$, and $N_{s} \text{ mod } 4 = 2$.}
 \label{table:naive-slopes}
\end{table}

For the two cases considered here, we tried various fits that attempted to incorporate subleading corrections.  We attempted fits with four or five free parameters.  These included corrections $\propto 1/N_{s}$, $1/N_{s}^{2}$, $1/\ln(N_{s})$ and $1/\ln^{2}(N_{s})$.  To judge the quality of the fits we compared the average relative error between fits,
\begin{equation}
	(\text{Relative Error})^2 = \frac{1}{N}\sum_{i = 1}^{N}\left(\frac{y_{i} - f(x_{i})}{y_{i}}\right)^{2}
\end{equation}
with $y_{i}$ the dependent data and $f(x_{i})$ the fitting function evaluated at the independent data.  This measure is convenient since the error is dimensionless; in addition, a $\chi^2$ measure of error would depend upon the unknown DMRG error bars and fitting with uncertainties in arbitrary units gives a relatively useless estimate of the fit quality.
The relative errors associated with the fits were never greater than $10^{-3}$, and never less than $10^{-7}$.   For systems with subsystems of size $l$, we considered a fit of the form
\begin{align}
\label{eq:cft-fit}
	S_{n}(N_{s}, l) &= A_{n} \ln \left\{ N_{s} \sin \left[  \frac{\pi l}{N_{s}} \right] \right\} + B \\
    &+ \frac{C}{N_{s}^{p_n}} \cos ( \pi l )
    \left| \sin \left[ \frac{\pi l}{N_{s}} \right] \right|^{-p_n} \nonumber \\
    &+ f_n(N_{s}, l) \nonumber
\end{align}
with $f_n$ a function to take into account additional corrections, and $A_{n}$, $B$, $C$, and $p_{n}$ are fit parameters.  However, we focused on data with $l=N_{s}/2$.  We found the best fit results by excluding data with $l < N_{s}/2$ and $l > N_{s} /2$ for small $N_{s}$ and at larger $N_{s}$ we found fits preferred data near $l \approx N_{s}/2$, resulting in data which resembled a ``fan with a handle'' (see \cite{myfirstprl}). 

For case 1 the best fits included corrections like $f \propto 1/N_{s}^{2}$ and $1/\ln^{2}(N_{s})$.  We found almost identical relative errors between corrections $1/N_{s}^{2}$ and $1/\ln^{2}(N_{s})$.  For case 2, the OBC data had the least error with corrections $\propto 1/N_{s}$ while the PBC data had the least error with corrections $\propto 1/N_{s}^{2}$.
For the oscillating term, the various $p_{n}$ are expected to follow special relations \cite{PhysRevLett.104.095701} (see below). 
For some fits, there were no oscillations present
and the fits drove $p_{n}$ very large. In these cases we replaced the $\pi N_{s}/2$ in the cosine by $\pi N_{s}$, so as to set it to unity by hand, and assumed a correction $\propto N_{s}^{-p_n}$.
The fits were done by nonlinear least-squares minimization.

\begin{figure}
	\includegraphics[width=0.5\textwidth]{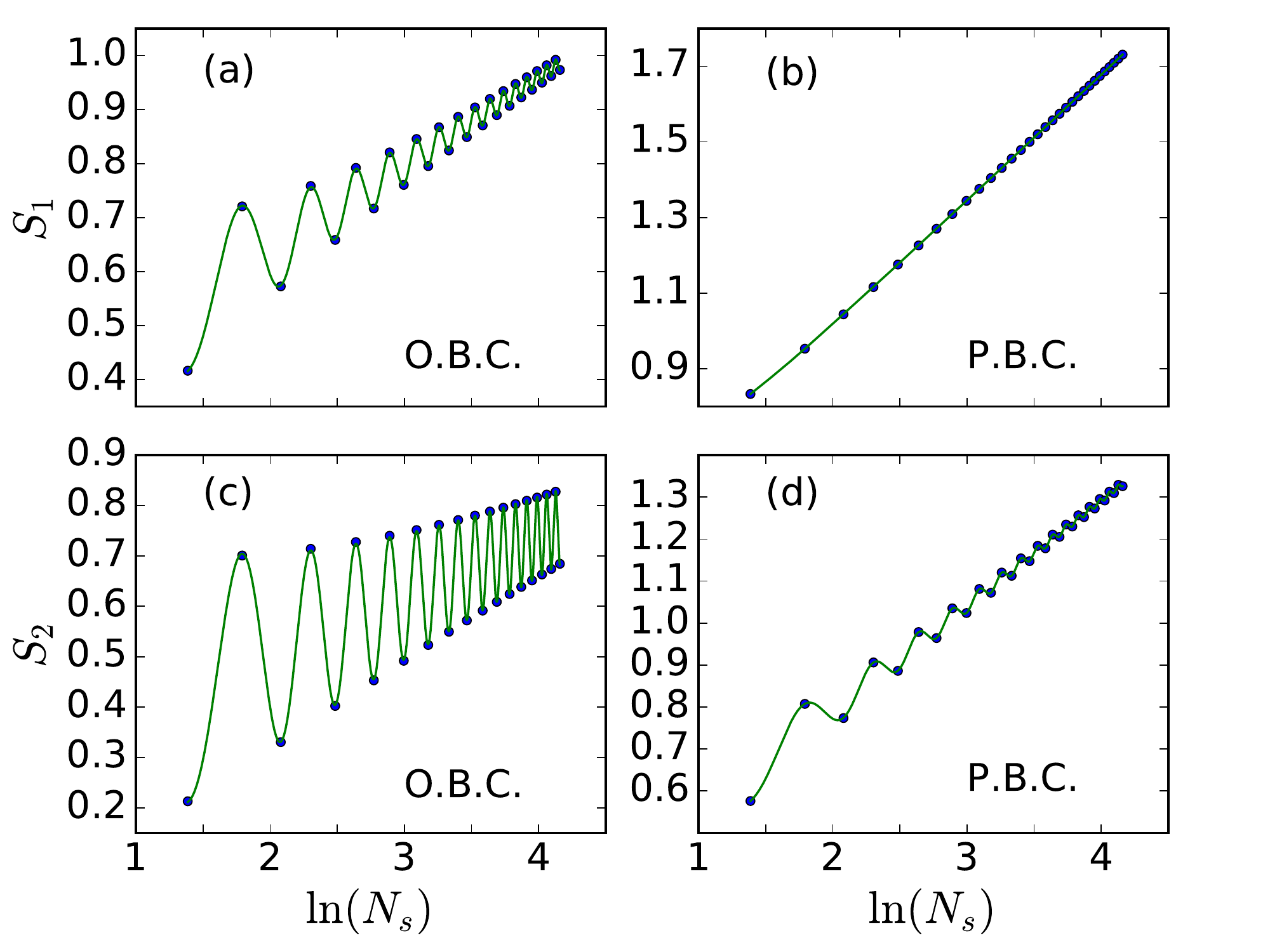}
	\caption{The first order and second order R\'{e}nyi entropy scaling with system size for $\beta_{s}\beta_{\tau} = 0.01$, $\mu\beta_{\tau} = 0.5$ in the time continuum limit calculated using DMRG.  (a), (b) The first order R\'{e}nyi entropy with OBC and PBC respectively.  (c), (d) The second order R\'{e}nyi entropy with OBC and PBC respectively.}
	\label{fig:cct1}
\end{figure}

\begin{figure}
	\includegraphics[width=0.5\textwidth]{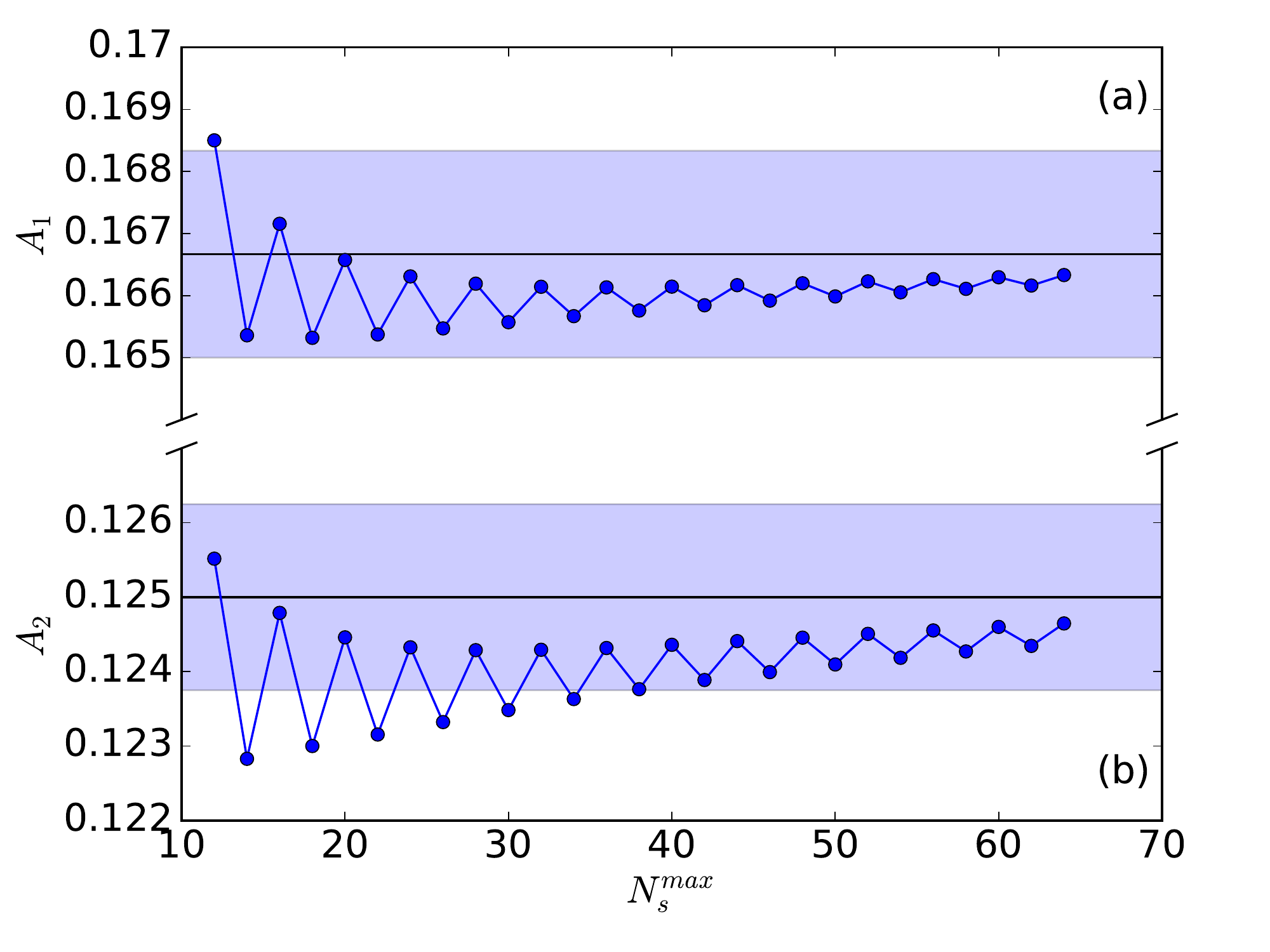}
	\caption{The $A_{n}$ for $\beta_{s}\beta_{\tau} = 0.01$, $\mu\beta_{\tau} = 0.5$ with OBC is plotted versus the maximal size of the lattice used to fit the data.  The horizontal line is the CFT prediction with $c=1$ with a region around representing a $\pm 1\%$ deviation. (a) The first order R\'{e}nyi entropy.  (b) The second order R\'{e}nyi entropy.}
	\label{fig:c1candcft}
\end{figure}

\begin{figure}
	\includegraphics[width=0.5\textwidth]{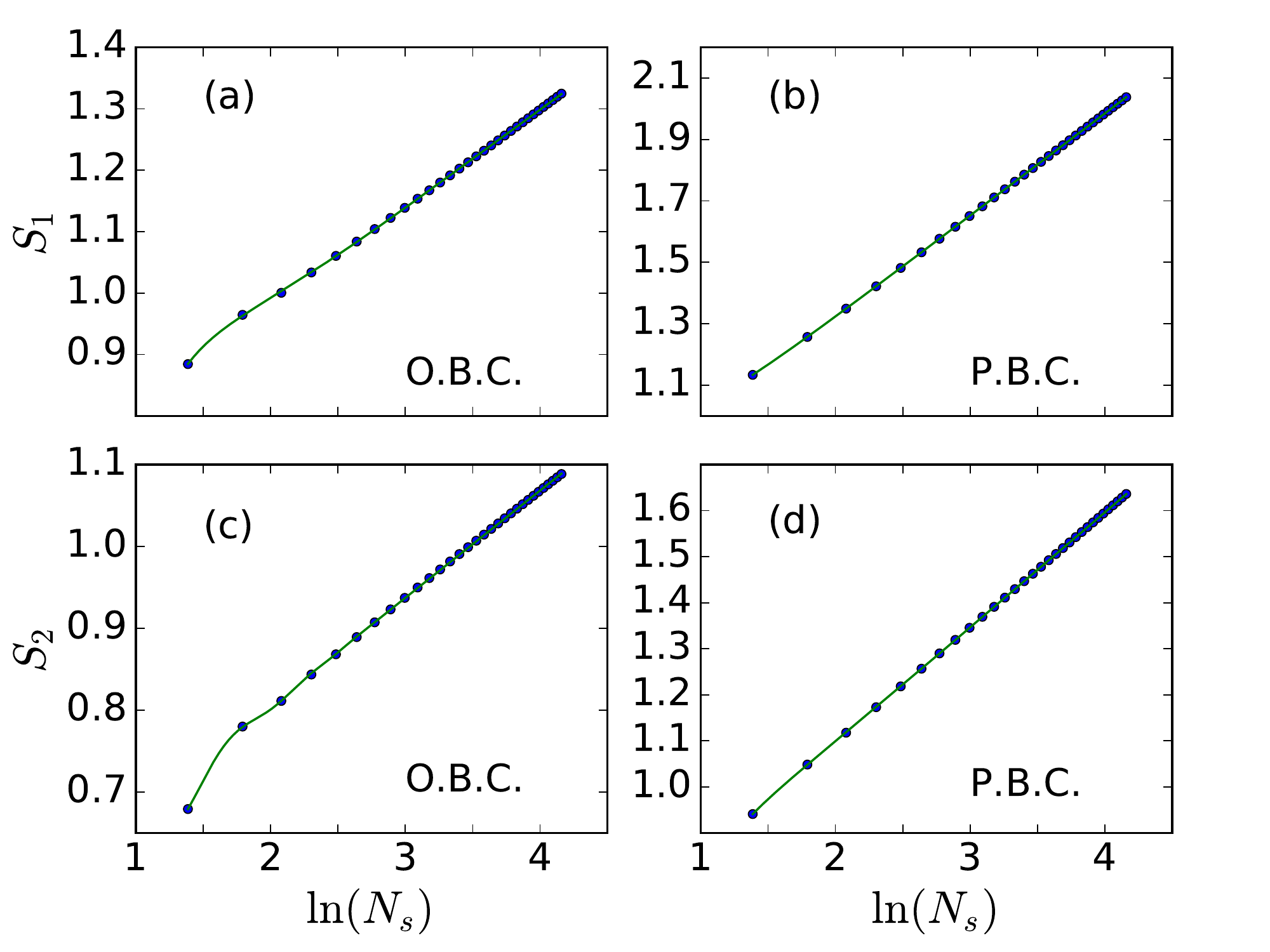}
	\caption{The first order and second order R\'{e}nyi entropy scaling with system size for $\beta_{s}\beta_{\tau} = 4$, $\mu\beta_{\tau} = 0$ in the time continuum limit calculated using DMRG.  (a), (b) The first order R\'{e}nyi entropy with OBC and PBC respectively.  (c), (d) The second order R\'{e}nyi entropy with OBC and PBC respectively.}
	\label{fig:cct2}
\end{figure}

The results are shown in Fig. \ref{fig:cct1} for case 1.
The values of the slopes, $A_{n}$, for both $S_{1}$ and $S_{2}$ are plotted in Fig. \ref{fig:c1candcft} with the slope value predicted from CFT surrounded by a band representing a 1\% deviation from the CFT value.
The values for $A_n$, and $p_{n}$ using all the data points up to $N_s=64$ are shown in Table \ref{table:dmrgfits}. Notice the good agreement with the predicted relations \cite{PhysRevLett.104.095701} $p_{1} ^{\text{OBC}}=2 p_{2}^{\text{OBC}}$, 
$p_{1}^{\text{PBC}} =2 p_{2}^{\text{PBC}}$ and 
$p_n^{\text{PBC}}=2p_n^{\text{OBC}}$.
\begin{table}
 \begin{tabular}{||c|c|c|c||}
     \hline 
     $S_{n}$ & $p_n$ & $A_n$ from fit & $A_n$ with $c=1$ \\
    \hline\hline
     $S_1$ PBC & 2.315  & 0.3338  & $0.\bar{3}$ \\ \hline
     $S_2$ PBC & 0.981  & 0.2525 & 0.25\\\hline   
    $S_1$ OBC & 0.901  & 0.1663 & $0.1\bar{6}$ \\ \hline
     $S_2$ OBC & 0.443  & 0.1246 & 0.125\\\hline   
\end{tabular}
 \caption{Values for $A_n$, and $p_{n}$ for $\beta_{s}\beta_{\tau} = 0.01$, $\mu \beta_{\tau} = 0.5$, from the least-squares fits to the DMRG data up to $N_s=64$ corresponding to
 Fig. \ref{fig:cct1}.}
 \label{table:dmrgfits}
\end{table}
The fit results for case 2 are shown in Fig. \ref{fig:cct2}.
As can be seen, the oscillations are very small, if at all, as compared to case 1.  Also in contrast, case 2 did not yield the special relationships between the $p_{n}$ exponents that did occur for case 1.  The $A_{n}$ values when fitting to all the data up to $N_{s} = 64$ are found in Table \ref{table:dmrgfits2}.
\begin{table}
 \begin{tabular}{||c|c|c|c||}
     \hline 
     $S_{n}$ & $A_n$ from fit & $A_n$ with $c=1$ \\
    \hline\hline
     $S_1$ PBC  & 0.3337  & $0.\bar{3}$ \\ \hline
     $S_2$ PBC  & 0.2500 & 0.25\\\hline   
    $S_1$ OBC  & 0.1654 & $0.1\bar{6}$ \\ \hline
     $S_2$ OBC  & 0.1278 & 0.125\\\hline   
  \end{tabular}
 \caption{Values for $A_n$ for $\beta_{s}\beta_{\tau} = 4$, $\mu\beta_{\tau} = 0$, from the least-squares fits to the DMRG data up to $N_s=64$ corresponding to
 Fig. \ref{fig:cct2}.}
 \label{table:dmrgfits2}
\end{table}
In both case 1 and case 2 the first order R\'{e}nyi (von Neumann) entropy with PBC possesses no oscillations, which is in agreement with what is known \cite{Fagotti-Calabrese:2011}.  These results suggest that both of these different regions of the phase diagram are conformal and approximately $c=1$.  If either of these two regimes could be quantum simulated and experimentally realized, it may be possible to measure the central charge.  We will briefly discuss the feasibility of this prospect in the next section.

\section{Prospect for Quantum Simulations}
\label{sec:qsim}

To better understand the possibility of quantum simulating the $O(2)$ model it is important to find a suitable condensed matter model to relate to.  We considered a single species Bose-Hubbard quantum Hamiltonian (Eq. \eqref{eq:BH-ham}) in a region of the phase diagram where the two models are essentially identical: $\tilde{\mu} \approx U/2 \gg J$, i.e. similar to case 1.  While the hopping parameter is very small compared to the on-site repulsion, the chain is only \emph{half-filled}, allowing the superfluid regime to be probed.

\begin{figure}
	\includegraphics[width=0.5\textwidth]{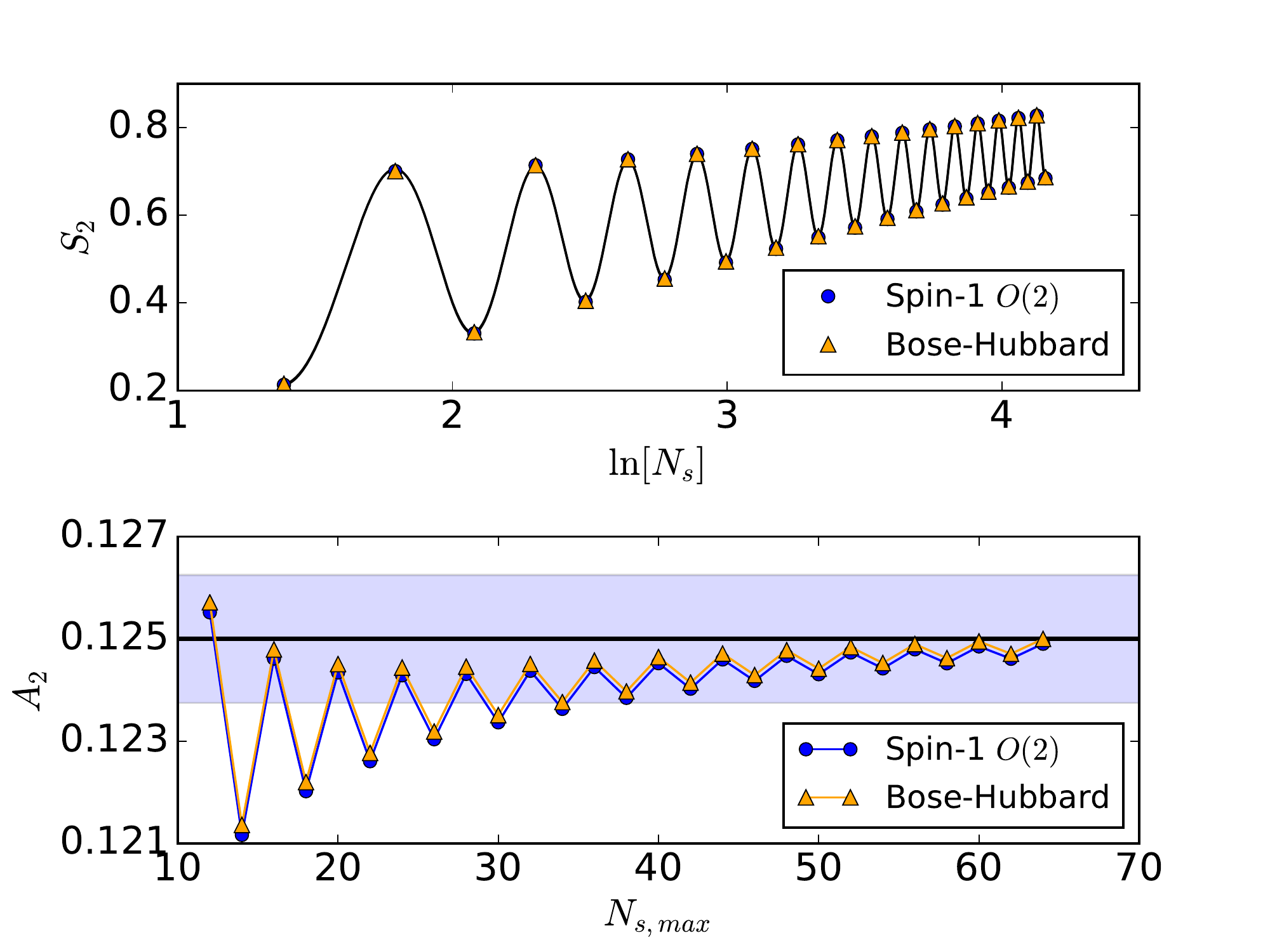}
    \caption{A comparison between the Bose-Hubbard model and the $O(2)$ model at half-filling and $J/U = 0.005$ with OBC.  (top) $S_{2}$ for the two models.  The data lie almost on top of each other, and there are two solid black lines fitting the two data sets which are essentially indistinguishable. (bottom) The value of $A_{2}$ versus the maximum size of the lattice used to extract the value of $A_{2}$.  The horizontal line represents the CFT prediction of $1/8$ with a band representing a $\pm 1 \%$ deviation.}
    \label{fig:BH-o2-comp}
\end{figure}

We considered the second order R\'{e}nyi entropy for the Bose-Hubbard model for $J/U = 0.005$ and $J/U = 0.1$ with OBC.  We did runs using DMRG across various system sizes such that $4 \leq N_{s} \leq 64$, and sub-system sizes such that $1 \leq l \leq N_{s}-1$.  To illuminate the legitimacy of the comparison between the two models we have plotted $S_{2}$ and $A_{2}$ for both the $O(2)$ model in the time-continuum limit for case 1, and the Bose-Hubbard model with $J/U = 0.005$ in Fig. \ref{fig:BH-o2-comp}.
As one can see in Fig. \ref{fig:BH-o2-comp} the BH model in this limit is almost identical to case 1 of the $O(2)$ model.  Changing $J/U$ to 0.1 increases the discrepancy, but the models continue to agree quantitatively well, especially for smaller volumes.  The exploratory fits and trials can be found in \cite{myfirstprl}.  A Bose-Hubbard model with small spatial volumes and $\mu = U/2 \gg J/U = 0.1$ appears as a potential candidate for quantum simulating the $O(2)$ model and experimentally measuring the central charge.  The possibility of measuring the central charge with cold atoms trapped in optical lattices was investigated in Ref. \cite{myfirstprl}.

\subsection*{Finite temperature effects}

For quantum simulation, while most of the calculations were done at $T=0$, finite temperature effects should be considered.  Here we take $k$, the Boltzmann constant, equal to unity.  While working on a two-dimensional Euclidean lattice, we must relate the temporal extent to the physical temperature.  This is done through the relation
\begin{equation}
	\frac{1}{T} = N_{\tau}a
\end{equation}
This relation can be derived with simple quantum statistical mechanics arguments.  To take the time continuum limit, we allowed $\beta_{\tau} \rightarrow \infty$ and $a \rightarrow 0$.  This allowed us to set the scale with the quantity $U \equiv 1/\beta_{\tau}a$.  With this definition we have
\begin{equation}
	\frac{1}{T} = \frac{N_{\tau}}{\beta_{\tau} U} = \frac{\beta_{s} N_{\tau}}{2J}.
\end{equation}
This relates a number of important quantities, for instance the spatial and temporal coupling on the Euclidean lattice to the physical temperature of a quantum Hamiltonian, as well as the number of temporal sites on the lattice.  In addition it relates the hopping parameter, $J$, and the on-site repulsion, $U$, to the physical temperature and lattice couplings.

\begin{figure}
	\includegraphics[width=8.6cm]{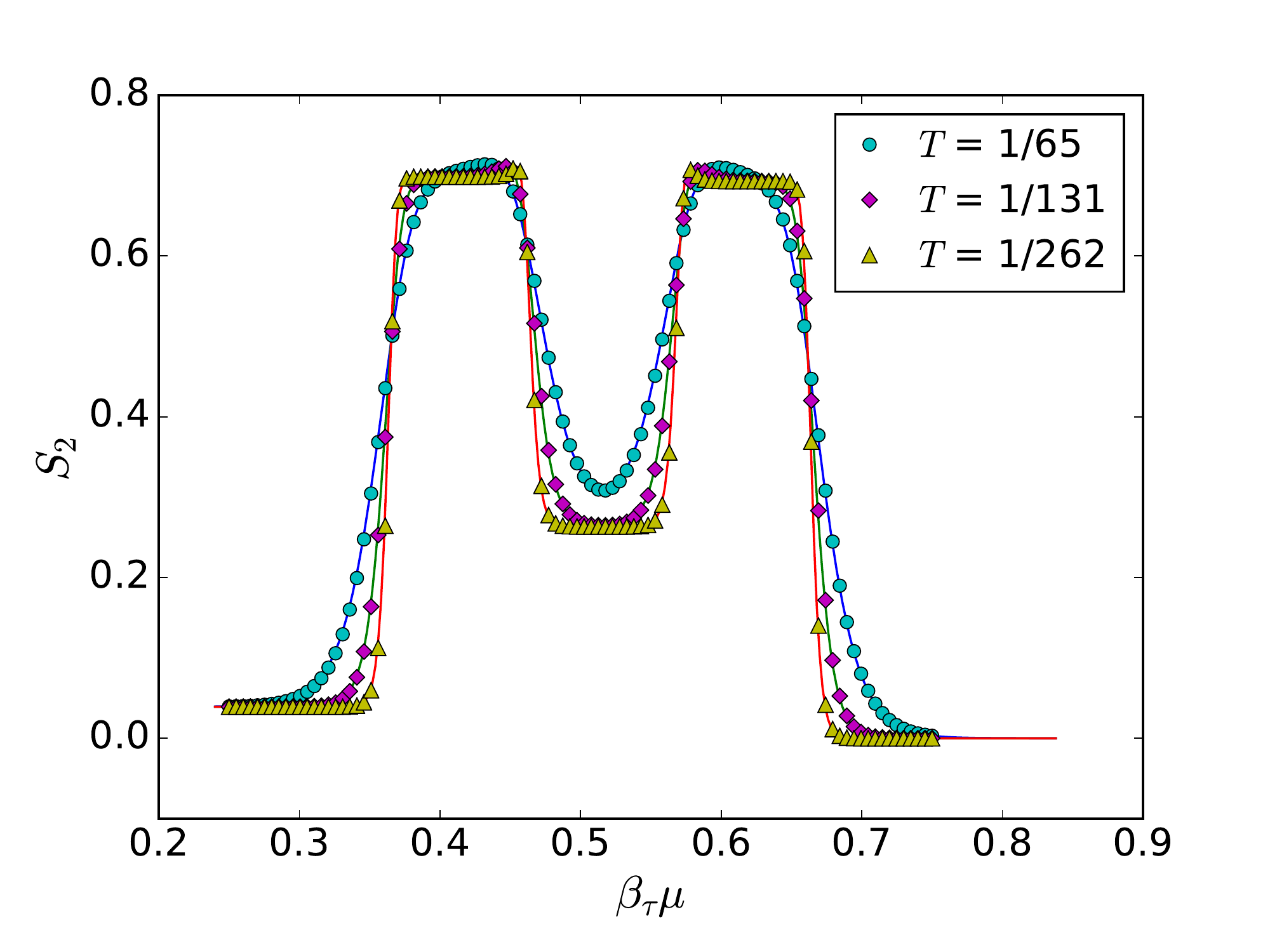}
    \caption{The second order R\'{e}nyi entropy as a function of the chemical potential, $\tilde{\mu} = \beta_{\tau}\mu$, for $N_{s} = 4$ with $J/U = 0.1$.  The solid lines are the DMRG data and the markers are the TRG data.  We see the effects of a finite temperature on $S_{2}$ is to smooth out the peaks and valleys.  As the temperature increases the thermal entropy dominates the R\'{e}nyi entropy.  The temperature, $T$, was calculated here with $U = 1$ and $\beta_{\tau} = 500$, and $D_{\text{bond}} = 201$ for the TRG.}
    \label{fig:finiteT-S2}
\end{figure}

To verify this relation between the classical picture of a two-dimensional lattice and its quantum counter-part in one less dimension we again compared time-continuum TRG results with DMRG results for finite-temperature.  For TRG we merely considered temporal lattice sizes which were not as great as before for the zero-temperature analysis and tuned the couplings for the time-continuum limit.  The DMRG analysis used a thermal density matrix (as opposed to using the ground state) to compute the R\'{e}nyi and von Neumann entanglement entropies \cite{PhysRevLett.93.207204,PhysRevLett.93.227205,PhysRevB.72.220401}.  In Fig. \ref{fig:finiteT-S2} one can see the agreement between the DMRG calculations and the TRG ones in the case of the $O(2)$ model.  This Figure also demonstrates the effect the thermal entropy has on the entanglement entropy as the temperature increases.  The peaks and valleys of the  entanglement entropy become smoothed out and while the boundaries remain at approximately $\ln(2)$, the half-filling valley increases.

For systems at half-filling, i.e. the central peak (valley) like in Fig. \ref{fig:entent_x4-8_sites} and \ref{fig:finiteT-S2}, the CC scaling can be well fit to a functional form
\begin{align}
	S_n(N_{s}) &= A_n \ln(N_{s}) + B \\
    &+ \frac{C}{N_{s}^{p_n}}\cos(\pi N_{s}/2) + E N_{s} + f(N_{s}),
    \nonumber
\end{align}
that is, adding a term linear in $N_{s}$ takes into account the finite temperature effects.  For fits to a general subsystem size at fixed $N_{s}$, we find a term linear in $l$ fits the data well,
\begin{align}
	\label{eq:xavier-T}
	S_{n}(N_{s}, l) &= A_{n} \ln \left\{ \frac{4(N_{s}+1)}{\pi} \sin \left[  \frac{\pi (2l+1)}{2(N_{s}+1)} \right] \right\} + B \\
    &+ \frac{C}{N_{s}^{p_{n}}} \sin \left[ \frac{\pi (2l+1)}{2} \right]
    \left| \sin \left[ \frac{\pi (2l+1)}{2(N_{s}+1)} \right] \right|^{-p_{n}} \nonumber \\
    &+ D l + f_{n}(N_{s}, l). \nonumber
\end{align}
This linear term can be used to subtract off finite-temperature effects \cite{Kaufman2016}, however we find additional corrections are necessary to maintain the original $T=0$ fit parameters.  Examples of fits done with these functional forms for various temperatures can be found in \cite{myfirstprl}.  

\section{Conclusion}
\label{sec:conclusion}

We have argued that finite-size scaling of the entanglement
entropy may provide a sensitive tool for identifying
conformal behavior in a system. It may complement
the techniques currently in use in the lattice gauge theory
community for studying models in the context of physics
Beyond the Standard Model. Such models are harder to study
on the lattice than QCD, because the running of the
coupling is slow and the relevant physics may be easily
masked by lattice artifacts (\textit{e.g.} finite lattice spacing,
finite volume).

We have calculated the R\'{e}nyi entropy for the classical $O(2)$ model in the isotropic coupling limit, as well as in the anisotropic coupling limit with a quantum Hamiltonian.  From fits to the R\'{e}nyi entropy we have estimated the central charge.  We found that this model can be mapped to a single-species Bose-Hubbard model in a particular region of the phase diagram and their R\'{e}nyi entropies are quantitatively similar allowing for the possibility of quantum simulating the $O(2)$ model and observing the Calabrese-Cardy scaling during simulation.  In addition we have considered finite-temperature effects on the R\'{e}nyi entropy, and found fitting functions which match the data for $S_{2}$ well, with scaling in $N_{s}$ and in subsystem size.  These additional fits involved including a term which is linear in either the subsystem size, $l$, or the system size, $N_{s}$.

It would be interesting to study the scaling of the R\'{e}nyi entropy of the $O(3)$ nonlinear sigma model with finite chemical potential in 1+1 dimensions.  This model is known to have asymptotic scaling in the continuum limit leading to a non-zero mass-gap, as well as meron (instanton) solutions due to the $O(2)$ sub-group.  The phase diagram in the time-continuum limit has a similar form \cite{PhysRevD.94.114503} to the $O(2)$ model considered here, and could be investigated in a similar fashion.

In addition it would be interesting to study the effects of a weak gauge-coupling to the $O(2)$ spins (as discussed in \cite{PhysRevD.19.1882,PhysRevD.92.076003}).  This would be scalar electrodynamics in a perturbative limit of weak gauge coupling.  Monitoring the entanglement entropy as one takes the limit of zero gauge coupling would give information about the symmetries for the phases of scalar electrodynamics, and the passage between the two models.

\begin{acknowledgments}
We thank M. C. Banuls, I. Bloch, I. Cirac, M. Greiner, A. Kaufman, G.  Ortiz, J. Osborn, H. Pichler and 
P. Zoller for  useful suggestions or comments, as well as Philipp Preiss for input on experimental aspects of quantum simulating with cold atoms on optical lattices. 
This research was supported in part  by the Department of Energy
under Award Numbers DOE grant DE-FG02-05ER41368, DE-SC0010114 and DE-FG02-91ER40664, the NSF under grant DMR-1411345 and by the Army Research Office of the Department of Defense under Award Number W911NF-13-1-0119.  
L.-P. Yang was supported by Natural Science Foundation for young scientists of China (Grants No.11304404) and Research Fund for the Central Universities(No. CQDXWL-2012-Z005).  Parts of the  numerical calculations were done  at the Argonne  Leadership  Computational Facilities. Y.M. thanks the Focus Group Physics with Effective Field Theories of the Institute for Advanced Study, Technische Universit{\"a}t M{\"u}nchen, and the workshop on ``Emergent properties of space-time" at CERN for hospitality while part of this work was carried out and the Institute for Nuclear Theory for motivating this work during the workshop ``Frontiers in Quantum Simulation with Cold Atoms".
\end{acknowledgments}


\end{document}